%% file: main.tex
\documentclass[ALICE,manyauthors]{cernphprep}

\newcommand{\pp}{p\kern-0.05em p}
\newcommand{\ppbar}{\mathrm{p}\kern-0.05em \overline{\mathrm{p}}}

\newcommand{\PbPb}{\ensuremath{\mbox{Pb--Pb}}}
\newcommand{\GeVc}{\ensuremath{\mathrm{GeV}\kern-0.05em/\kern-0.02em c}}
\newcommand{\sqrts}{\ensuremath{\sqrt{s_{\mathrm{NN}}}}}
\newcommand{\pT}{\ensuremath{p_{\mathrm{T}}}}

\newcommand{\etajet}{\ensuremath{\eta_{\mathrm{jet}}}}

\newcommand{\pTchjet}{\ensuremath{p_{\mathrm{T}}^{\mathrm{ch\; jet}}}}

\mathchardef\mhyphen="2D
\newcommand{\antikT}{\ensuremath{\mathrm{anti\mhyphen}k_\mathrm{T}}}

\newcommand{\zcut}{\ensuremath{z_{\mathrm{cut}}}}

\newcommand{\DeltaR}{\ensuremath{\Delta R_{\mathrm{axis}}}}

\usepackage[comma,square,numbers,sort&compress]{natbib}
\usepackage{hyperref}
\usepackage{lineno}
\usepackage{amsmath}
\usepackage{bigints}  
\usepackage{afterpage}  
\usepackage{multirow}
\usepackage[dvipsnames]{xcolor}
\usepackage[Symbol]{upgreek}
\usepackage{bm}
\usepackage[T1]{fontenc}
\usepackage{orcidlink} 

\begin{document}%

\begin{titlepage}
\PHyear{2023}       
\PHnumber{046}      
\PHdate{20 March}  

\title{Measurement of the angle between jet axes in \PbPb{} collisions at $\mathbf{\sqrt{{\textit s}_{\text NN}} \bm{=} 5.02}  \ \mathbf{TeV}$}

\ShortTitle{Measurement of angle between jet axes in $\sqrts = 5.02$ TeV \PbPb{} collisions}   

\Collaboration{ALICE Collaboration\thanks{See Appendix~\ref{app:collab} for the list of collaboration members}}
\ShortAuthor{ALICE Collaboration} 

\begin{abstract}

This paper presents the first measurement of the angle between different jet axes (denoted as \DeltaR) in
\PbPb{} collisions. The measurement is carried out in the 0--10\% most-central events at $\sqrts = 5.02$ TeV.
Jets are assembled by clustering charged particles at midrapidity using the \antikT{} algorithm with resolution parameters $R=0.2$ and $0.4$
and transverse momenta in the intervals $40 < \pTchjet < 140 \ \GeVc$ and $80 < \pTchjet < 140 \ \GeVc$, respectively. Measurements at these low transverse momenta enhance the sensitivity to quark--gluon plasma (QGP) effects. A comparison to models implementing various mechanisms of jet energy loss in the QGP suggests that the observed narrowing of the \PbPb{} distribution relative to \pp{} can be explained if quark-initiated jets are more likely to emerge from the medium than gluon-initiated jets.
These new measurements disfavor intra-jet \pT{} broadening as implemented in a simple model calculation with the BDMPS formalism for energy loss in the QGP.
The comparison of \PbPb ~and pp collisions shows sensitivity to the angular scale at which the QGP can resolve two independent splittings, favoring mechanisms that incorporate incoherent energy loss.

\end{abstract}
\end{titlepage}
\setcounter{page}{2}

\input{Ch1_Introduction}

\input{Ch2_ExperimentalSetup}
\input{Ch3_DataAnalysis}

\input{Ch4_SystematicUncertainties}

\input{Ch5_Results}

\input{Ch6_Conclusion} 

\newenvironment{acknowledgement}{\relax}{\relax}
\begin{acknowledgement}
\section*{Acknowledgements}
We gratefully acknowledge Felix Ringer, Daniel Pablos, and the JETSCAPE collaboration for providing theoretical predictions. We also acknowledge Xin-Nian Wang and Feng Yuan for fruitful discussions.
\input{fa_2023-03-08_Opt_C.tex}
\end{acknowledgement}

\section*{Data availability statement}
This manuscript has associated data in the HEPData repository~\cite {HEPData}.

\bibliographystyle{utphys}   
\bibliography{main}

\newpage
\appendix
\section{The ALICE Collaboration}
\label{app:collab}
\input{2023-03-08-Alice_Authorlist_2023-03-08_Opt_C.tex}  
\end{document}

%% file: Ch1_Introduction.tex
\begin{figure}[h]
    \centering
    \includegraphics[width=0.9\textwidth]{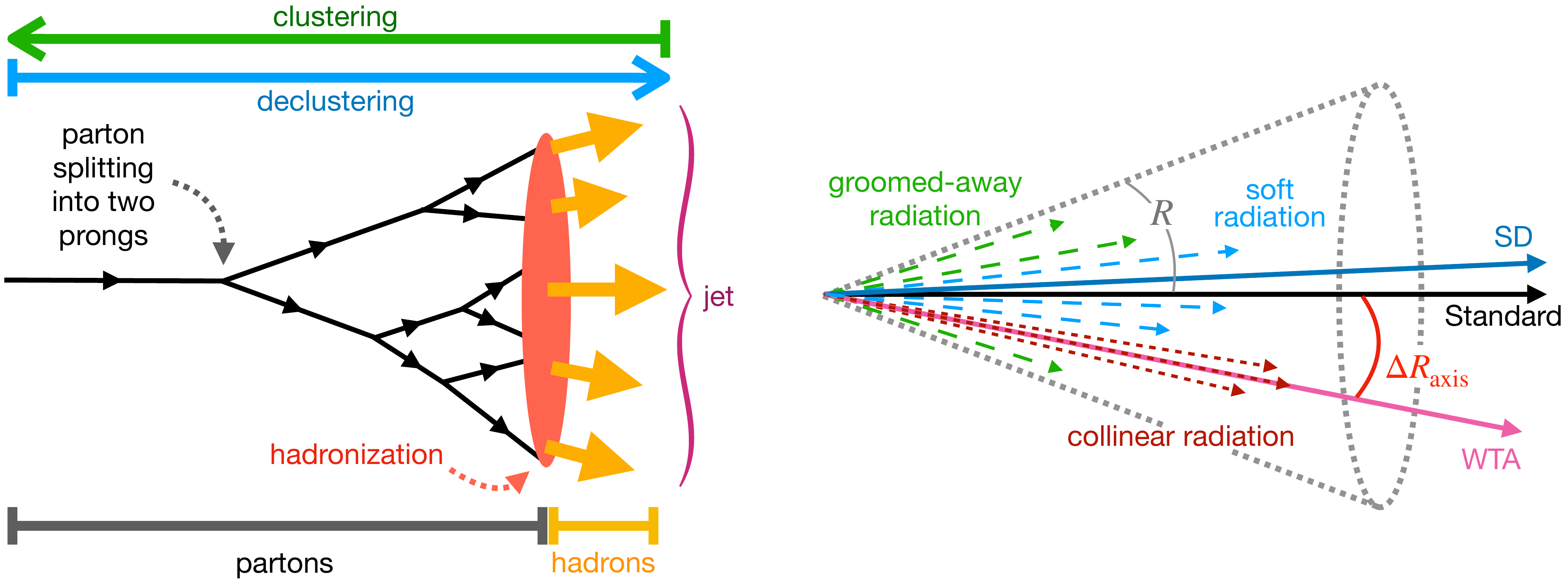}
    \caption{Left: schematic of how energetic partons split or fragment into lower-energy prongs or branches of partons down to a scale in which hadrons emerge. An algorithm can then be run over final-state hadrons to cluster them into jets. An inverse declustering algorithm can be used to recursively explore the jet substructure. Right: schematic showing that different jet axes (solid arrows) can be selected from a jet to construct the $\DeltaR$ observable. The Standard axis (represented by a black arrow) is constrained by all the jet constituents (represented by dashed arrows). The SD axis (represented by a dark blue arrow) is determined by the particles left in the jet after grooming (blue and dark red arrows). Finally, the WTA axis (represented by a pink arrow) tends to be aligned with the most energetic constituent.}
    \label{fig:cartoon}
\end{figure}

The strongly interacting, deconfined state of matter of our universe a few microseconds after the Big Bang
can be recreated in ultrarelativistic heavy-ion collisions at the Large Hadron Collider (LHC)~\cite{Busza:2018rrf,ALICE:2022wpn}. Studying the microscopic structure of this quark--gluon plasma (QGP) can elucidate the physics of many-body strongly coupled systems~\cite{Jacak:2012dx}.
Because droplets of QGP created at colliders are small and short-lived, studying interactions within those droplets requires auto-generated probes. In this study, jets of energetic hadrons, which arise from high-momentum-transfer (hard) collisions between quarks and gluons (partons) in the incoming nuclei, are used. These hard-scattered particles produce showers of secondary partons, which give rise to jets of multiple correlated hadrons (see Fig.~\ref{fig:cartoon}, left).

When traversing the QGP, jets  interact with the medium, lose energy, and are modified relative to their structure in elementary (e.g., ee or pp) collisions~\cite{Acharya:2019jyg, ATLAS:2018gwx, CMS:2021vui, STAR:2020xiv, ALICE:2021obz, ALICE:2018dxf, ATLAS:2018bvp, ATLAS:2019dsv, ATLAS:2023hso, CMS:2014jjt, CMS:2018mqn, CMS:2017qlm, CMS:2018fof, ALICE:2015mdb, STAR:2017hhs}. 
Thus, measuring the substructure of jets can provide sensitivity to different scales within the jet modification process. 
The direction of a jet in rapidity ($y$) and azimuth ($\varphi$), referred to as the jet axis, 
is affected by transverse modification of the partonic shower, with varying sensitivity to soft radiation, depending on how the axis is defined~\cite{Cal:2019gxa}.

This paper presents the first measurement in heavy-ion collisions of the angle between pairs of jet axes,  $\DeltaR$ (see Fig.~\ref{fig:cartoon}, right). This observable shows how well the QGP resolves the angular structure of the parton shower~\cite{Cal:2019gxa} and is defined as
\begin{equation}
    \Delta R_{\rm axis} \equiv \sqrt{(y_{\rm axis, \, 1}-y_{\rm axis, \, 2})^2+(\varphi_{\rm axis, \, 1}-\varphi_{\rm axis, \, 2})^2},
\end{equation}
where $y_{{\rm axis}, \, i}$ and $\varphi_{{\rm axis}, \, i}$ are the rapidity and azimuthal coordinates of the axes, respectively.
The axes considered in this analysis are:

\begin{itemize}

\item ``Standard'': coordinates in $y$--$\varphi$ of the 4-vector resulting from clustering jets with the \antikT{} algorithm~\cite{Cacciari:2008gp} using the $E$ recombination scheme (two branches combined into one by adding their 4-momenta). In this case, all jet constituents including soft (low-transverse-momentum) and hard radiation contribute to the axis.

\item 
``Soft Drop'' (SD): coordinates in $y$--$\varphi$ of jet after soft wide-angle radiation is removed using 
the Soft Drop grooming procedure~\cite{Larkoski:2014wba} with the Cambridge--Aachen (C/A) reclustering algorithm~\cite{Dokshitzer:1997in,Wobisch:1998wt}. The procedure stops when a splitting is found that satisfies the condition
\begin{equation}
    \frac{\mathrm{min}(p_{\rm T,1},p_{\rm T,2})}{p_{\rm T,1}+p_{\rm T,2}}>\zcut\left(\frac{\Delta R_{1,2}}{R}\right)^{\beta},
    \label{eq:sd}
\end{equation}
where $p_{{\rm T},i}$ are the transverse momenta of each of the two branches in the splitting, $\Delta R_{1,2}$ is the distance between them in the $y$--$\varphi$ plane, $R$ is the jet resolution parameter, and \zcut{} and $\beta$ are 
user-set parameters.

\item ``Winner-Takes-All'' (WTA): coordinates in $y$--$\varphi$ of the jet after reclustering its constituents using the C/A algorithm and the WTA \pT{} recombination scheme~\cite{Bertolini:2013iqa}, which combines two jet branches into one by giving the merged branch the total \pT{} of the two sub-branches and the direction of the hardest sub-branch.
The WTA axis is often consistent with the direction of the hardest constituent of the jet and is insensitive to soft radiation at the leading power of the observable~\cite{Cal:2019gxa}.

\end{itemize}

Measurements of $\DeltaR$ are perturbatively calculable, but also sensitive to medium-induced effects; we examine sensitivity to 
such effects by studying jets of different \pT{} and radius parameter, $R$. 
An increase in the relative fraction of quark- vs. gluon-initiated jets due to medium-induced radiative energy loss can narrow the angular distribution, while multiple scattering of the hard-scattered parton off medium partons can broaden it~\cite {Ringer:2019rfk}.
Reference~\citenum{ALICE:2021obz} reported a net narrowing of the distribution in \PbPb{} collisions, quantitatively consistent with the former.
This can be expected if the broadening was primarily in the wide-angle radiation that grooming removes from the jet. This is tested by looking for broadening effects in an ungroomed angular observable, such as the angle between the Standard and WTA axes.
Increasing the grooming intensity
and simultaneously varying the jet resolution parameter $R$ can uncover medium-induced 
effects. Comparing experimental results with 
multiple models 
can help disentangle medium response and medium-induced large-angle radiation.

The medium resolution length, $L_{\rm res}$, is a property of the QGP proportional to the inverse of its local temperature~\cite{Hulcher:2017cpt}. It characterizes the minimum separation distance at which two color charges within the same jet interact independently with the QGP. Certain jet substructure observables are sensitive to this scale~\cite{Casalderrey-Solana:2019ubu}. In this analysis, the sensitivity of $\DeltaR$ to $L_{\rm res}$ is investigated.

The measured $\DeltaR$ distributions in \pp{} collisions at $\sqrt{s} = 5.02 \ {\rm TeV}$ reported in Ref.~\citenum{ALICE:2022rdg} are used as the reference for QGP-induced effects reported here.
Perturbative QCD calculations~\cite{Cal:2019gxa} were found to agree with the measurements in \pp{} collisions within the experimental uncertainties, demonstrating good theoretical control of the observables.

%% file: Ch2_ExperimentalSetup.tex
Data were collected in 2018 from \PbPb{} collisions at a center-of-mass energy per nucleon--nucleon pair of $\sqrts = 5.02$ TeV with the ALICE detector~\cite{ALICE:2022wpn}. A detailed description of the detector and its performance can be found in Refs.~\citenum{ALICE:2008ngc} and \citenum{ALICE:2014sbx}. A trigger based on the charged-particle multiplicity
in the forward region~\cite{ALICE:2013axi} was used to select events in the 0--10\% centrality class~\cite{ALICE:2013hur,ALICE:2018tvk}. The events were required to have a reconstructed primary vertex within $\pm 10$ cm of the nominal interaction point in the longitudinal direction. Beam-induced background events were removed using two Zero Degree Calorimeters and pileup effects were removed by discarding events that contained multiple reconstructed vertices~\cite{Acharya:2019jyg}. These selection criteria yield a data sample with 92 million events, corresponding to an integrated luminosity of $\approx 0.12 \ {\rm nb}^{-1}$.
 
Charged-particle tracks reconstructed in the Inner Tracking System (ITS)~\cite{ALICE:2010tia} and Time Projection Chamber (TPC)~\cite{Alme:2010ke} used in the analysis were required to have a transverse momentum $\pT > 0.15 \ \GeVc$ and a pseudorapidity $|\eta|<0.9$. For details, see the supplemental materials~\cite{JA:sup}, including references [2, 5] therein.
The detector performance was estimated by propagating simulated events through a GEANT~3~\cite{Brun:1119728} model of the ALICE detector.
The transverse-momentum resolution $\sigma(\pT)/\pT$ increases linearly from $\approx 1\%$ at $\pT=1 \ \GeVc$ to $\approx4\%$ at $\pT=50 \ \GeVc$~\cite{ALICE:2014sbx}. The tracking efficiency rapidly increases from $\approx 65\%$ at $\pT=0.15 \ \GeVc$ to $\approx 85\%$ at $\pT=1 \ \GeVc$, and remains above $\approx 75\%$ at higher \pT.

%% file: Ch3_DataAnalysis.tex
The large uncorrelated background from the underlying event (UE)~\cite{ALICE:2012nbx} in high-multiplicity heavy-ion collisions is subtracted before jet finding using the event-wide constituent subtraction procedure from Refs.~\citenum{Berta:2014eza} and~\citenum{Berta:2019hnj}. The parameters used are $\alpha=0$, $A_{\rm g} = 0.01$, and $\Delta R^{\rm max}= 0.1$ (0.25) when the track sample is used to cluster jets of $R=0.2$ (0.4).

Tracks included in jet clustering are assigned the charged-pion mass. Charged-particle jets are clustered with the \antikT{} algorithm, the $E$ recombination scheme, and resolution parameters $R=0.2$ and 0.4, using FastJet~\cite{Cacciari:2011ma}. Jets containing a track with $\pT>100 \ \GeVc$ and jets with their Standard axis not pointing inside the fiducial volume of the TPC ($|\etajet|<0.9-R$) are discarded
to minimize contamination by fake or partially reconstructed jets

The measured distributions are unfolded to correct for detector effects as well as residual UE fluctuations using an iterative Bayesian procedure~\cite{DAgostini,DAgostini:2010hil}. The procedure uses a 4D response matrix (RM) that captures the correspondence between \DeltaR{} and jet transverse momentum (\pTchjet) at the generator and detector levels. The generator-level component of the RM is modeled using generated PYTHIA 8~\cite{Sjostrand:2014zea} Monash 2013~\cite{Skands:2014pea} events. The detector-level component of the RM is modeled by propagating those events through a GEANT~3~\cite{Brun:1119728} model of the ALICE detector and embedding the smeared events into measured \PbPb{} events.

%% file: Ch4_SystematicUncertainties.tex
Systematic uncertainties 
arise from the tracking inefficiency, unfolding procedure, unfolding model dependence, UE subtraction procedure, and residual effects from combinatorial jets.
The tracking-inefficiency uncertainty is determined by repeating the analysis with an RM based on a track sample with 4\% of all tracks randomly rejected. This number corresponds to the uncertainty in the tracking efficiency, determined by varying the track selection parameters and ITS--TPC matching criteria.

The unfolding-procedure uncertainty is determined by extending the lower and upper bounds of the \pTchjet{} spectrum by 5 \GeVc, changing the detector-level \DeltaR{} binning, changing the number of unfolding iterations, and scaling the prior distribution. The four categories of variations are carried out individually and combined into an unfolding uncertainty using their standard deviation. 
The uncertainty from the model dependence in the unfolding is determined by repeating the analysis with response matrices based on different event generators including Herwig~7~\cite{Gieseke:2012ft} and JEWEL~\cite{Zapp:2013vla} with and without the inclusion of medium partons recoiling after interactions with the jet.

Potential bias introduced by the event-wide UE subtraction~\cite{Berta:2014eza,Berta:2019hnj} is quantified as the maximum difference between the nominal result and the results obtained by repeating the procedure with parameters $\Delta R^{\rm max}= 0.05$ (0.1), which lead to an under-subtraction, and $\Delta R^{\rm max}= 0.5$ (0.7), which lead to an over-subtraction of the background, for jets of $R=0.2$ (0.4).
Deviations between the generator-level and unfolded distributions due to residual combinatorial jets are determined with a closure test. PYTHIA~8 events are embedded into background events with no jets and analyzed identically to the measured data. Deviations between the unfolded and PYTHIA~8 generator-level spectra are assigned as the uncertainty.

The systematic uncertainties are added in quadrature. The total systematic uncertainty is in the range of 2--61\% depending on \DeltaR{} and \pTchjet. 
The large variation comes primarily from the background subtraction procedure at low \pTchjet{} (where background effects are largest). It is small at low values of \DeltaR, and smoothly grows in the tails of the distributions. 
The supplemental materials~\cite{JA:sup} provide details.

%% file: Ch5_Results.tex
The measured distributions are normalized to the total number of jets in a given \pTchjet{} interval, $N$, which is equivalent to reporting the self-normalized cross section:
\begin{equation}
\frac{1}{N(\pTchjet)} \frac{{\rm d}N}{{\rm d}\DeltaR} \left(\pTchjet\right)
\equiv
\frac{1}{\sigma(\pTchjet)} \frac{{\rm d}\sigma}{{\rm d}\DeltaR} \left(\pTchjet\right).
\end{equation}
This normalization choice causes the \PbPb/pp{} ratios to go above unity at low $\DeltaR$. In an absolute cross section measurement, the  ratio would be below unity as a result of jet suppression in heavy-ion collisions. 
The suppression factor was measured in Ref.~\citenum{ALICE:RAA23}.
Thus, only the shape and not the absolute scale of the \DeltaR{} spectra is reported.

\begin{figure}
    \centering
    \includegraphics[width=0.497\textwidth,trim={0.8cm 0 0.3cm 0.7cm},clip]{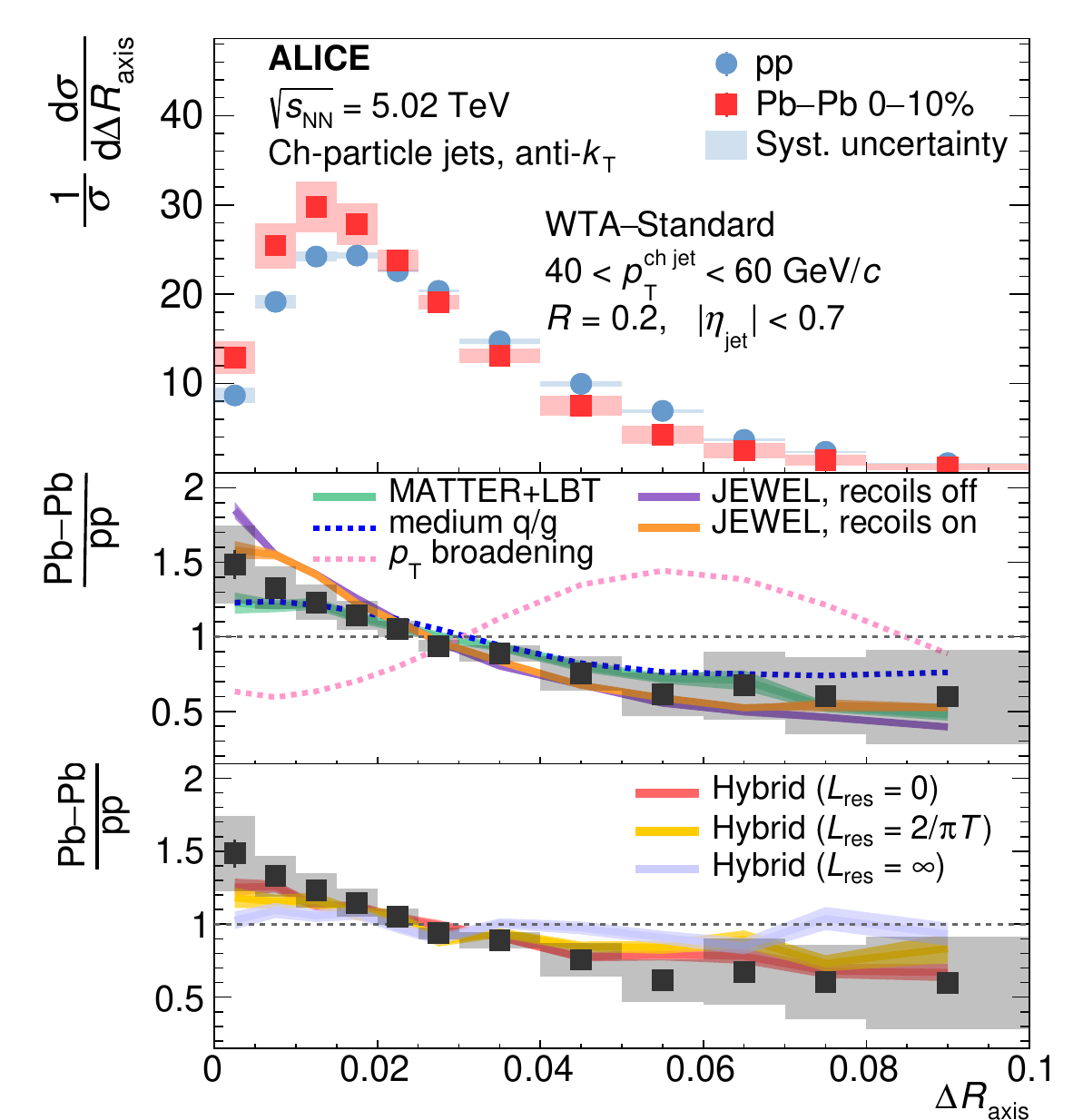}
    \includegraphics[width=0.497\textwidth,trim={0.8cm 0 0.3cm 0.7cm},clip]{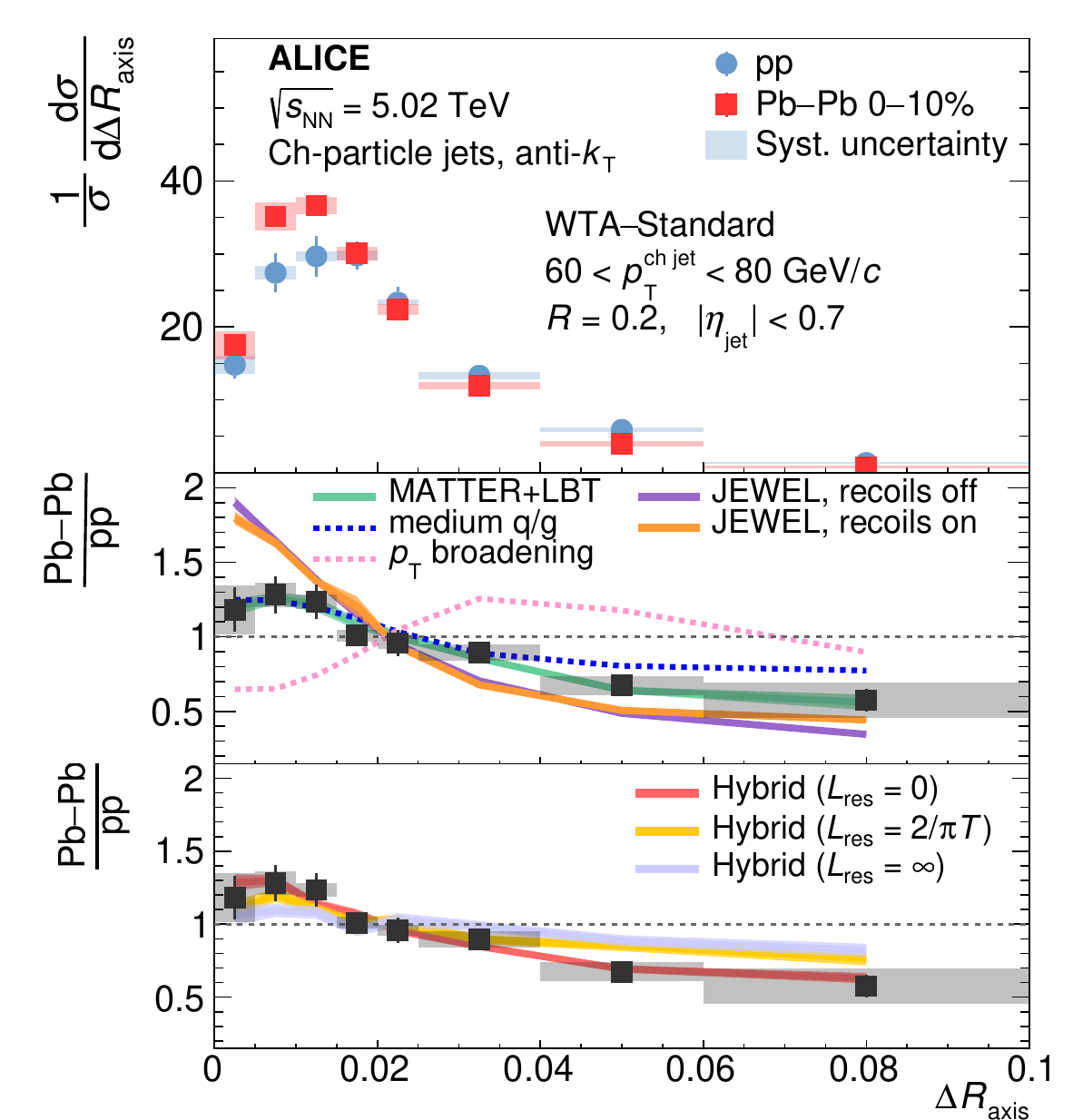}
    \caption{Top panels: fully corrected \PbPb{} and \pp{} $\DeltaR^{\rm WTA-Standard}$ distributions in the \pTchjet{} intervals $[40,60]$ (left), and $[60,80]$ (right) \GeVc{} for jets of $R=0.2$.
    The \pp{} baseline is taken from Ref.~\citenum{ALICE:2022rdg}.
    Central and bottom panels: measured \PbPb/\pp{} ratio in black, as well as predictions from a selection of jet quenching models.
    }
    \label{fig:R02_WTA_Standard}
\end{figure}

The measured WTA--Standard distributions are shown in Fig.~\ref{fig:R02_WTA_Standard} for $R=0.2$ jets in the \pTchjet{} ranges $[40,60]$ and $[60,80]$ \GeVc. The top panels show \DeltaR{} spectra from \PbPb{} and \pp{} collisions.
The vertical error bars represent the statistical uncertainties and the rectangles represent the total systematic uncertainties.
The central and bottom panels show the \PbPb/\pp{} ratio, with all uncertainties assumed uncorrelated and added in quadrature when calculating the ratio.
The equivalent results for other \pTchjet{} ranges, jet resolution parameters, and grooming settings are included in the supplemental materials~\cite{JA:sup}.

The data are compared with several jet quenching models. These models have different implementations of the microscopic properties of the medium, its evolution, and the jet--medium interaction.
The \textbf{JEWEL} event generator~\cite{Zapp:2013vla}
models the medium with a boost-invariant longitudinally expanding ideal quark--gluon gas~\cite{Zapp:2012ak}. Parameters from Ref.~\citenum{KunnawalkamElayavalli:2017hxo}, which are adequate for the kinematics of this measurement, are used.
The medium partons recoiling after interacting with jet constituents can be discarded from the event (recoils off) or
allowed to hadronize together with the jet (recoils on).
\textbf{``MATTER+LBT''} 
is from the JETSCAPE event generator~\cite{Putschke:2019yrg,JETSCAPE:2022jer}, implementing an in-medium parton shower with interactions of high- (low-) virtuality partons with the medium described by the MATTER~\cite{Majumder_2013} (Linear Boltzmann Transport~\cite{LBT}) model.
The curve labeled \textbf{``medium q/g''} corresponds to a phenomenological model in which the only difference between the \PbPb{} and \pp{} results comes from a modification of the fraction of quarks and gluons that initiate the jets~\cite{Qiu:2019sfj,Qiu:2020pus}, highlighting that these two jet populations lose energy differently in the medium.
The \textbf{``$\bm{p_{\rm T}}$ broadening''} calculation adds to the previous model a \pT{} broadening caused by incoherent multiple scatterings with the medium partons~\cite{Ringer:2019rfk} following the he Baier-Dokshitzer-Mueller-Peigne-Schiff (BDMPS) approach~\cite{Baier:1996kr,Baier:1996sk,Baier:1998kq}.
This calculation uses a mean square momentum transfer coefficient between the jet and medium constituents $\langle \hat q L \rangle=5$ GeV$^2$.
\textbf{``Hybrid''} is from the hybrid model~\cite{Casalderrey-Solana:2014bpa}, which describes the weakly coupled jet showering process using the Dokshitzer-Gribov-Lipatov-Altarelli-Parisi (DGLAP) formalism and the strongly coupled interaction between the jet constituents and medium partons via holographic calculations of energy loss based on AdS/CFT.
The case $L_{\rm res}=0$ corresponds to fully incoherent energy loss, where the medium is able to resolve the splitting immediately after the parton fragments, resulting in higher energy loss~\cite{Hulcher:2017cpt}.
The case $L_{\rm res}=\infty$ corresponds to fully coherent energy loss, where the medium is not able to resolve splittings, 
and the energy loss is lower.
The intermediate case $L_{\rm res}=2/\uppi T$, where $T$ is the medium local temperature, is also included. 
All ``Hybrid'' curves include medium-response effects from Ref.~\citenum{Casalderrey-Solana:2016jvj}.


Fig.~\ref{fig:R02_WTA_Standard} shows that the $R=0.2$ distributions in \PbPb{} are narrower than those measured in \pp{} collisions, as was also observed in Ref.~\cite{ALICE:2021obz}. 
The narrowing may be due to a selection bias, as jets selected using \pTchjet{} 
(rather than the initial momentum before the jet-medium interactions, which is not accessible experimentally) tend to come from the surface and lose little energy in the QGP~\cite{Brewer:2021hmh}. 
The data are 
also consistent with a quark-enhanced mixture of jets surviving the interaction with the medium relative to the vacuum case. This is expected since 
gluon-initiated jets couple more strongly to partons of the QGP and thus lose more energy~\cite{Qiu:2019sfj,Spousta:2015fca,ALICE:2021obz}.
The MATTER+LBT model generally describes the data for all considered kinematical regions. 

\begin{figure}
    \centering
    \includegraphics[width=0.497\textwidth,trim={0.88cm 0 0.3cm 0.62cm},clip]{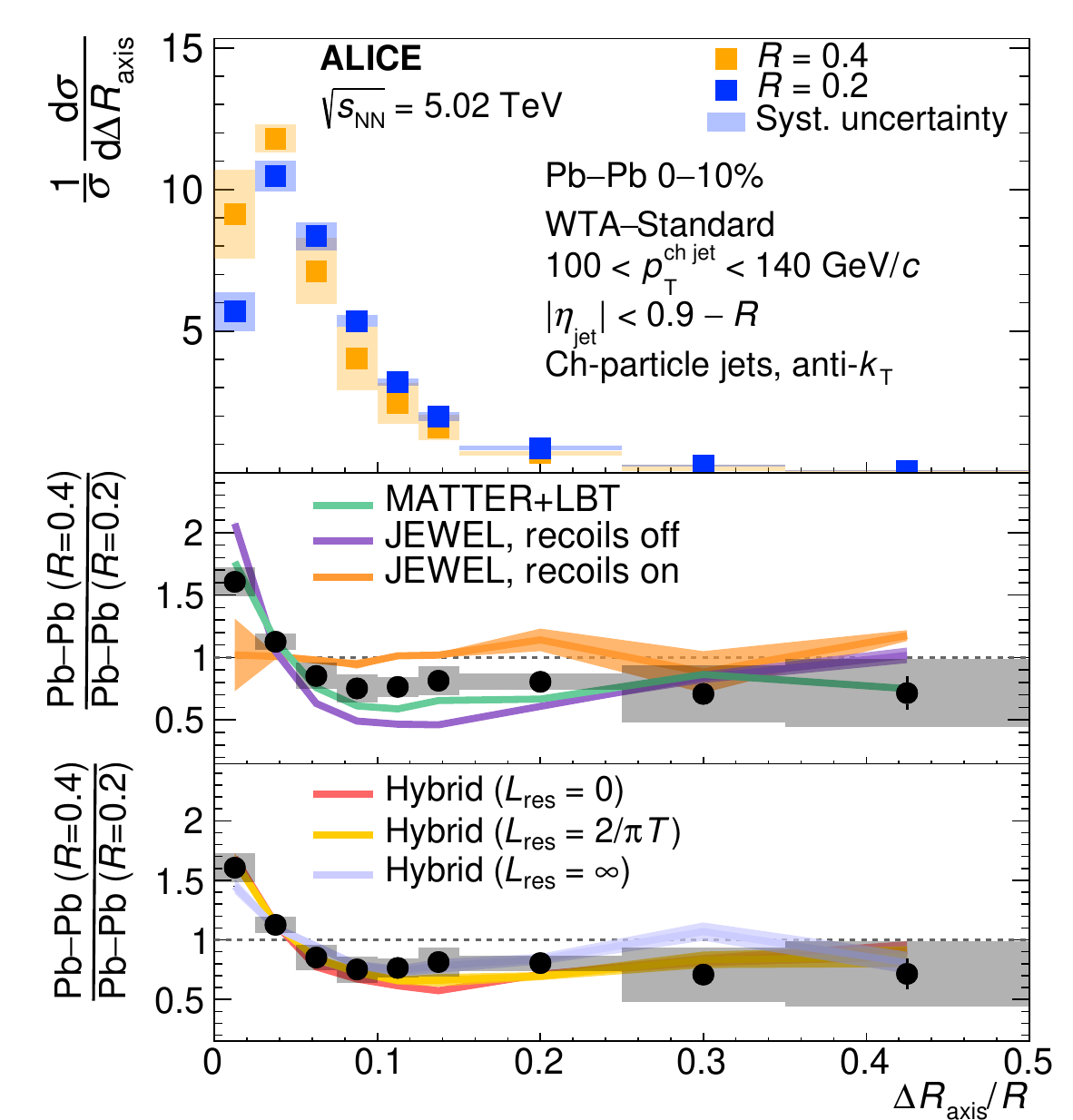}
    \includegraphics[width=0.497\textwidth,trim={0.88cm 0 0.3cm 0.62cm},clip]{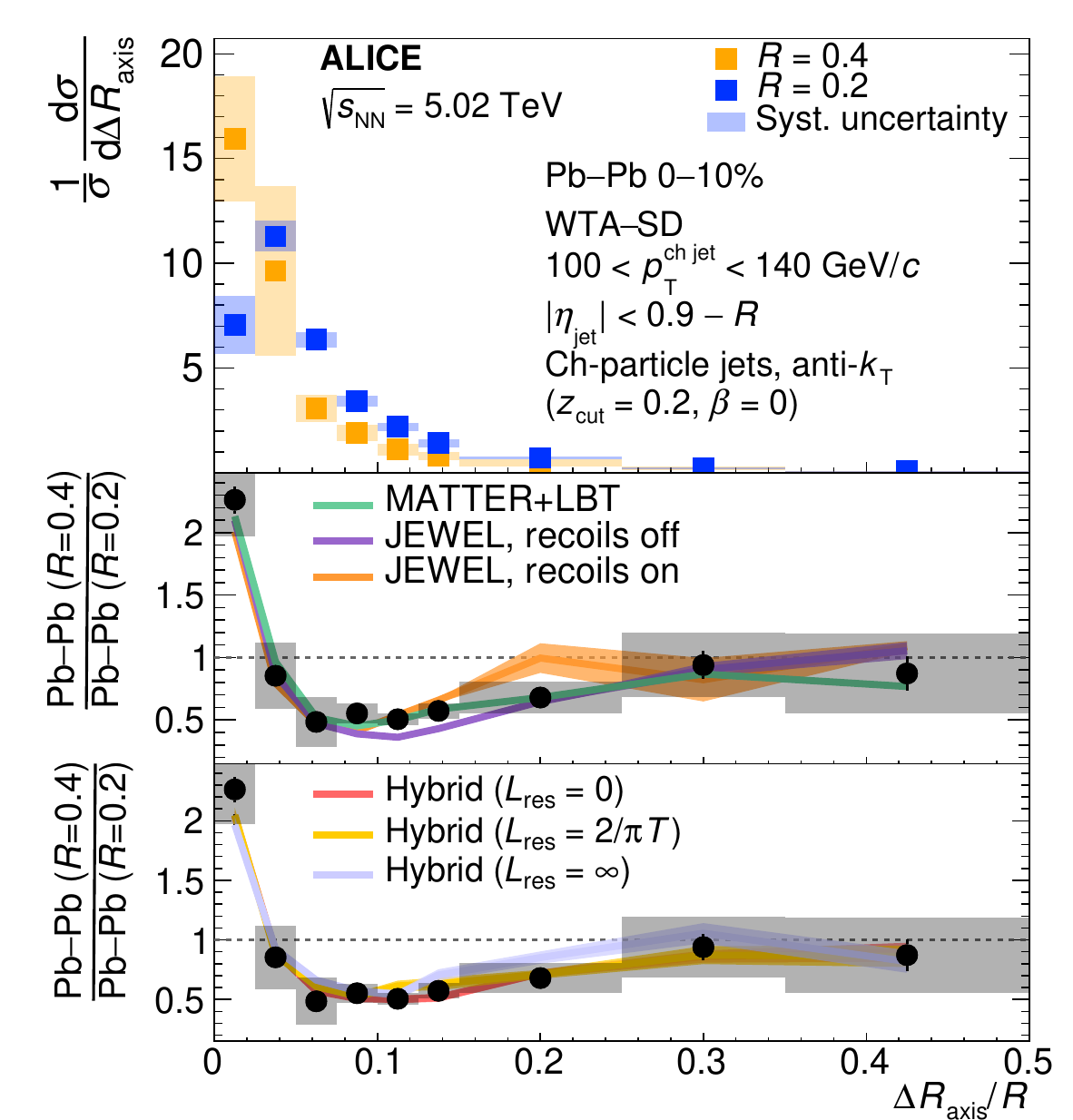}
    \caption{Top panels: fully corrected \PbPb{} $\DeltaR^{\rm WTA-Standard}/R$ (left), and $\DeltaR^{\rm WTA-SD}/R$ with $(\zcut=0.2,\beta=0)$ (right) distributions in the \pTchjet{} interval $[100,140]$ \GeVc.
    Central and bottom panels: measured $\PbPb (R=0.4)/\PbPb(R=0.2)$ ratio in black, as well as predictions from a selection of jet quenching models.
    }
    \label{fig:100_140}
\end{figure}

Fig.~\ref{fig:100_140} compares jet axis difference distributions for 100-140 GeV/c jets with  $R=0.2$ and $R=0.4$. We note that $R=0.4$ jets have significantly larger systematic uncertainty due to larger background corrections; the systematics partially cancel in the ratio for $R=0.2$ and $R=0.4$. Consequently, model predictions are compared to this ratio.
In contrast with $R=0.2$, for larger $R$ there are significant differences in JEWEL predictions with and without recoils for $R=0.4$ jets. This is expected, as the medium response 
populates larger angles away from the Standard jet axis. 
The Hybrid model does not predict a significant effect on the jet axis from the medium response; 
see Ref.
~\cite{JA:sup} for details. Comparison to other models in which the medium response can be turned on and off (e.g. the jet wake in Co-LBT~\cite{Chen:2017zte,Zhao:2021vmu}) could help clarify the sensitivity of $\DeltaR$ to the properties and magnitude of this effect.
Narrowing of the distributions relative to those measured in \pp{} collisions for jets of $R=0.2$ is qualitatively predicted by all models except for ``$p_{\rm T}$ broadening'', which is calculated in a relatively simple formulation with a single value of $\hat{q}$. This tends to disagree with the hypothesis that intra-jet \pT{} broadening manifests in the groomed radiation in previous measurements~\cite{ALICE:2021obz}.

The data are 
closest to incoherent energy loss of jets in the Hybrid model ($L_{\rm res}=0$);
the fully coherent case ($L_{\rm res}=\infty$) predicts significantly less narrowing.
Incoherent energy loss better reproduces the data than the intermediate case of $L_{\rm res}=2/\uppi T$
even though holographic calculations of the Debye screening length indicate that the intermediate case 
is 
closer to the true medium resolution scale~\cite{Hulcher:2017cpt}.

The precision of the data at higher \pTchjet{} is limited by the integrated luminosity of the \pp{} dataset, which only allows for an extraction of the \PbPb/\pp{} results up to 100 \GeVc. However, in heavy-ion collisions, larger \pTchjet{} can be studied through the ratio of $(R=0.4)/(R=0.2)$ spectra, where the systematic uncertainties partially cancel. This kinematic range is also important for future comparisons with results from other experiments.
The ratio is reported as a function of the jet angular scale $\DeltaR/R$ that, by construction, is independent of $R$~\cite{Cal:2019gxa}. Figure~\ref{fig:100_140} shows the ratio for $\pTchjet \in [100,140]$  \GeVc{} in ungroomed and groomed jets, together with theoretical predictions. The left side shows that the Standard-WTA jet axis difference is narrower (closer to zero) for R=0.4 than R=0.2 jets. This implies, perhaps surprisingly, that jets including more wide angle radiation have their standard jet axis closer to the WTA. In groomed jets, the jet axis is even more aligned with the WTA. It should be noted, however, that the grooming here uses $\beta$=0, which means that there is no angular weight and any soft branch can be groomed away.

The ungroomed WTA--Standard result is challenging to models, as can be seen
in Fig.~\ref{fig:100_140} (left).
Grooming improves the agreement between the models and the data (see Fig.~\ref{fig:100_140} (right)), suggesting that the ungroomed result captures significantly more of the non-perturbative aspects of jet quenching. The $(R=0.4)/(R=0.2)$ ratio of WTA--Standard jet axis is not very sensitive to $L_{\rm res}$ either with or without grooming.
We have run PYTHIA8 MC to compare the ratios R=0.2/R=0.4 and we find that the WTA-Standard difference distribution does not change with the jet resolution parameter. In this, PYTHIA agrees with the measurement in PbPb.

\begin{figure}
    \centering
    \includegraphics[width=0.497\textwidth,trim={0.88cm 0 0.3cm 0.62cm},clip]{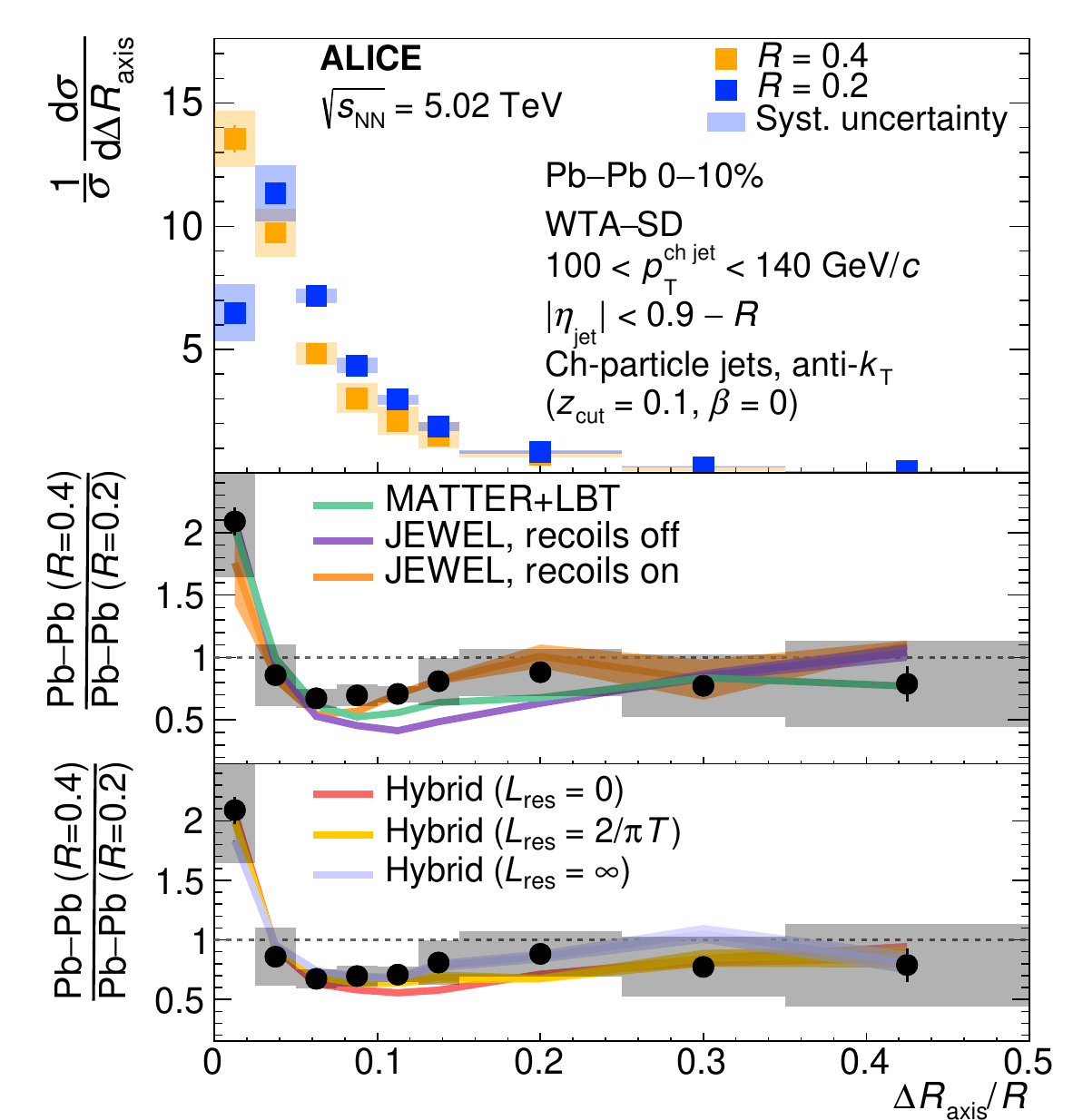}
    \caption{Fully corrected WTA--SD ($\zcut=0.1,\beta=0$) distributions in the $100<\pTchjet<140 \ {\rm GeV}/c$ interval.}
    \label{fig:grooming01}
\end{figure}

Fig.~\ref{fig:grooming01} shows an intermediate grooming setting for 100-140 \GeVc{} jets. 
One can see by
comparing $\DeltaR^{\rm WTA-Standard}$ to $\DeltaR^{\rm WTA-SD}$ with different grooming settings (Fig.~\ref{fig:grooming01} vs. Fig.~\ref{fig:100_140} right)
that once the softest branches are removed, further grooming does not substantially
impact the jet axis.
This was also seen in \pp{} collisions, and shows that the Standard and SD axes are more aligned with each other than either of them is with the WTA axis. 
While the Standard axis is influenced by the soft wide-angle radiation present in R=0.4 jets, grooming away more of the soft radiation does not further change the SD axis. This shows that once the softest particles are groomed away, the jet axis is dominated by the hard radiation. The models generally reproduce the intermediate grooming result, with the exception of JEWEL with recoils off. Stronger grooming improves the agreement, though some tension with the data remains. This reflects the importance of including medium response when calculating jet substructure in \PbPb{} collisions.
PYTHIA studies agree with the general trend, but do not quantitatively reproduce the ratios. Given the medium effects, this is not surprising.

%% file: Ch6_Conclusion.tex
In summary, the first measurement of the angle between the WTA and Standard or Soft Drop groomed charged-particle jet axes in \PbPb{} collisions is reported. The measurement was carried out using the 0--10\% most-central collisions at $\sqrts = 5.02$ TeV with the ALICE detector and for $\pTchjet \in [40,140]$ ([80,140]) \GeVc{} for jets of $R=0.2$ ($0.4$).

Narrowing of the \DeltaR{} distribution relative to the vacuum case is found. The narrowing may be explained if 
the \PbPb{} distribution is dominated by quark-initiated jets, since gluon-initiated jets are expected to interact more with the medium. This is consistent with the prediction from the ``medium q/g'' model. 
A selection bias causing apparent narrowing of the jet core should be reflected in all of the models which include energy loss. However, the distributions for ungroomed jets are not equally well described by the models, suggesting that the fate of large angle prongs in the medium is important. The measurement disfavors 
intra-jet \pT{} broadening, as implemented in 
Ref.~\citenum{Ringer:2019rfk} utilizing a single value of $\hat{q}$.
The \DeltaR{} observable 
is sensitive to the medium resolution length, favoring incoherent energy loss of jets, where the medium resolves a splitting immediately after it happens, regardless of the angular separation between the prongs. 

For jets of 100-140 GeV $p_T$, R=0.4 jet axes are more aligned with the WTA axis than those with R=0.2. Grooming with $\beta =0$ enhances this alignment. While grooming decreases the effects of medium response, fully reproducing both groomed and ungroomed jet axis differences requires including medium response in the models. 
Future measurements of \DeltaR{} in photon- or Z-boson-tagged jets could provide a way to experimentally vary the fraction of quark- vs. gluon-initiated jets and explore the relationship of the WTA axis to the initial parton direction.

%% file: fa_2023-03-08_Opt_C.tex

The ALICE Collaboration would like to thank all its engineers and technicians for their invaluable contributions to the construction of the experiment and the CERN accelerator teams for the outstanding performance of the LHC complex.
The ALICE Collaboration gratefully acknowledges the resources and support provided by all Grid centres and the Worldwide LHC Computing Grid (WLCG) collaboration.
The ALICE Collaboration acknowledges the following funding agencies for their support in building and running the ALICE detector:
A. I. Alikhanyan National Science Laboratory (Yerevan Physics Institute) Foundation (ANSL), State Committee of Science and World Federation of Scientists (WFS), Armenia;
Austrian Academy of Sciences, Austrian Science Fund (FWF): [M 2467-N36] and Nationalstiftung f\"{u}r Forschung, Technologie und Entwicklung, Austria;
Ministry of Communications and High Technologies, National Nuclear Research Center, Azerbaijan;
Rede Nacional de Física de Altas Energias (Renafae), Financiadora de Estudos e Projetos (Finep), Funda\c{c}\~{a}o de Amparo \`{a} Pesquisa do Estado de S\~{a}o Paulo (FAPESP) and Universidade Federal do Rio Grande do Sul (UFRGS), Brazil;
Bulgarian Ministry of Education and Science, within the National Roadmap for Research Infrastructures 2020-2027 (object CERN), Bulgaria;
Ministry of Education of China (MOEC) , Ministry of Science \& Technology of China (MSTC) and National Natural Science Foundation of China (NSFC), China;
Ministry of Science and Education and Croatian Science Foundation, Croatia;
Centro de Aplicaciones Tecnol\'{o}gicas y Desarrollo Nuclear (CEADEN), Cubaenerg\'{\i}a, Cuba;
Ministry of Education, Youth and Sports of the Czech Republic, Czech Republic;
The Danish Council for Independent Research | Natural Sciences, the VILLUM FONDEN and Danish National Research Foundation (DNRF), Denmark;
Helsinki Institute of Physics (HIP), Finland;
Commissariat \`{a} l'Energie Atomique (CEA) and Institut National de Physique Nucl\'{e}aire et de Physique des Particules (IN2P3) and Centre National de la Recherche Scientifique (CNRS), France;
Bundesministerium f\"{u}r Forschung, Technologie und Raumfahrt (BMFTR) and GSI Helmholtzzentrum f\"{u}r Schwerionenforschung GmbH, Germany;
General Secretariat for Research and Technology, Ministry of Education, Research and Religions, Greece;
National Research, Development and Innovation Office, Hungary;
Department of Atomic Energy Government of India (DAE), Department of Science and Technology, Government of India (DST), University Grants Commission, Government of India (UGC) and Council of Scientific and Industrial Research (CSIR), India;
National Research and Innovation Agency - BRIN, Indonesia;
Istituto Nazionale di Fisica Nucleare (INFN), Italy;
Japanese Ministry of Education, Culture, Sports, Science and Technology (MEXT) and Japan Society for the Promotion of Science (JSPS) KAKENHI, Japan;
Consejo Nacional de Ciencia (CONACYT) y Tecnolog\'{i}a, through Fondo de Cooperaci\'{o}n Internacional en Ciencia y Tecnolog\'{i}a (FONCICYT) and Direcci\'{o}n General de Asuntos del Personal Academico (DGAPA), Mexico;
Nederlandse Organisatie voor Wetenschappelijk Onderzoek (NWO), Netherlands;
The Research Council of Norway, Norway;
Commission on Science and Technology for Sustainable Development in the South (COMSATS), Pakistan;
Pontificia Universidad Cat\'{o}lica del Per\'{u}, Peru;
Ministry of Science and Higher Education, National Science Centre and WUT ID-UB, Poland;
Korea Institute of Science and Technology Information and National Research Foundation of Korea (NRF), Republic of Korea;
Ministry of Education and Scientific Research, Institute of Atomic Physics, Ministry of Research and Innovation and Institute of Atomic Physics and Universitatea Nationala de Stiinta si Tehnologie Politehnica Bucuresti, Romania;
Ministerstvo skolstva, vyskumu, vyvoja a mladeze SR, Slovakia;
National Research Foundation of South Africa, South Africa;
Swedish Research Council (VR) and Knut \& Alice Wallenberg Foundation (KAW), Sweden;
European Organization for Nuclear Research, Switzerland;
Suranaree University of Technology (SUT), National Science and Technology Development Agency (NSTDA) and National Science, Research and Innovation Fund (NSRF via PMU-B B05F650021), Thailand;
Turkish Energy, Nuclear and Mineral Research Agency (TENMAK), Turkey;
National Academy of  Sciences of Ukraine, Ukraine;
Science and Technology Facilities Council (STFC), United Kingdom;
National Science Foundation of the United States of America (NSF) and United States Department of Energy, Office of Nuclear Physics (DOE NP), United States of America.
In addition, individual groups or members have received support from:
Czech Science Foundation (grant no. 23-07499S), Czech Republic;
European Research Council (grant nos. 950692), European Union;
Strong 2020 - Horizon 2020 (grant nos. 824093), European Union;
ICSC - Centro Nazionale di Ricerca in High Performance Computing, Big Data and Quantum Computing, European Union - NextGenerationEU;
Academy of Finland (Center of Excellence in Quark Matter) (grant nos. 346327, 346328), Finland;
Deutsche Forschungs Gemeinschaft (DFG, German Research Foundation) ``Neutrinos and Dark Matter in Astro- and Particle Physics'' (grant no. SFB 1258), Germany;
Programa de Apoyos para la Superaci\'{o}n del Personal Acad\'{e}mico, UNAM, Mexico.

%% file: 2023-03-08-Alice_Authorlist_2023-03-08_Opt_C.tex
\begin{flushleft} 
\small

S.~Acharya\,\orcidlink{0000-0002-9213-5329}\,$^{\rm 126}$, 
D.~Adamov\'{a}\,\orcidlink{0000-0002-0504-7428}\,$^{\rm 87}$, 
A.~Adler$^{\rm 70}$, 
G.~Aglieri Rinella\,\orcidlink{0000-0002-9611-3696}\,$^{\rm 32}$, 
M.~Agnello\,\orcidlink{0000-0002-0760-5075}\,$^{\rm 29}$, 
N.~Agrawal\,\orcidlink{0000-0003-0348-9836}\,$^{\rm 51}$, 
Z.~Ahammed\,\orcidlink{0000-0001-5241-7412}\,$^{\rm 134}$, 
S.~Ahmad\,\orcidlink{0000-0003-0497-5705}\,$^{\rm 15}$, 
S.U.~Ahn\,\orcidlink{0000-0001-8847-489X}\,$^{\rm 71}$, 
I.~Ahuja\,\orcidlink{0000-0002-4417-1392}\,$^{\rm 36}$
% ...

\section*{Affiliation Notes}

$^{\rm I}$ Deceased\\
$^{\rm II}$ Also at: Max-Planck-Institut fur Physik, Munich, Germany\\
$^{\rm III}$ Also at: Italian National Agency for New Technologies, Energy and Sustainable Economic Development (ENEA), Bologna, Italy\\
$^{\rm IV}$ Also at: Dipartimento DET del Politecnico di Torino, Turin, Italy\\
$^{\rm V}$ Also at: Lehigh University, United States\\
$^{\rm VI}$ Also at: Department of Applied Physics, Aligarh Muslim University, Aligarh, India\\
$^{\rm VII}$ Also at: Institute of Theoretical Physics, University of Wroclaw, Poland\\

\section*{Collaboration Institutes}

$^{1}$ A.I. Alikhanyan National Science Laboratory (Yerevan Physics Institute) Foundation, Yerevan, Armenia\\
$^{2}$ AGH University of Krakow, Cracow, Poland\\
$^{3}$ Bogolyubov Institute for Theoretical Physics, National Academy of Sciences of Ukraine, Kyiv, Ukraine\\
$^{4}$ Bose Institute, Department of Physics  and Centre for Astroparticle Physics and Space Science (CAPSS), Kolkata, India\\
$^{5}$ California Polytechnic State University, San Luis Obispo, California, United States\\
$^{6}$ Central China Normal University, Wuhan, China\\
$^{7}$ Centro de Aplicaciones Tecnol\'{o}gicas y Desarrollo Nuclear (CEADEN), Havana, Cuba\\
$^{8}$ Centro de Investigaci\'{o}n y de Estudios Avanzados (CINVESTAV), Mexico City and M\'{e}rida, Mexico\\
$^{9}$ Chicago State University, Chicago, Illinois, United States\\
$^{10}$ China Nuclear Data Center, China Institute of Atomic Energy, Beijing, China\\
$^{11}$ Chungbuk National University, Cheongju, Republic of Korea\\
$^{12}$ Comenius University Bratislava, Faculty of Mathematics, Physics and Informatics, Bratislava, Slovak Republic\\
$^{13}$ COMSATS University Islamabad, Islamabad, Pakistan\\
$^{14}$ Creighton University, Omaha, Nebraska, United States\\
$^{15}$ Department of Physics, Aligarh Muslim University, Aligarh, India\\
$^{16}$ Department of Physics, Pusan National University, Pusan, Republic of Korea\\
$^{17}$ Department of Physics, Sejong University, Seoul, Republic of Korea\\
$^{18}$ Department of Physics, University of California, Berkeley, California, United States\\
$^{19}$ Department of Physics, University of Oslo, Oslo, Norway\\
$^{20}$ Department of Physics and Technology, University of Bergen, Bergen, Norway\\
$^{21}$ Dipartimento di Fisica, Universit\`{a} di Pavia, Pavia, Italy\\
$^{22}$ Dipartimento di Fisica dell'Universit\`{a} and Sezione INFN, Cagliari, Italy\\
$^{23}$ Dipartimento di Fisica dell'Universit\`{a} and Sezione INFN, Trieste, Italy\\
$^{24}$ Dipartimento di Fisica dell'Universit\`{a} and Sezione INFN, Turin, Italy\\
$^{25}$ Dipartimento di Fisica e Astronomia dell'Universit\`{a} and Sezione INFN, Bologna, Italy\\
$^{26}$ Dipartimento di Fisica e Astronomia dell'Universit\`{a} and Sezione INFN, Catania, Italy\\
$^{27}$ Dipartimento di Fisica e Astronomia dell'Universit\`{a} and Sezione INFN, Padova, Italy\\
$^{28}$ Dipartimento di Fisica `E.R.~Caianiello' dell'Universit\`{a} and Gruppo Collegato INFN, Salerno, Italy\\
$^{29}$ Dipartimento DISAT del Politecnico and Sezione INFN, Turin, Italy\\
$^{30}$ Dipartimento di Scienze MIFT, Universit\`{a} di Messina, Messina, Italy\\
$^{31}$ Dipartimento Interateneo di Fisica `M.~Merlin' and Sezione INFN, Bari, Italy\\
$^{32}$ European Organization for Nuclear Research (CERN), Geneva, Switzerland\\
$^{33}$ Faculty of Electrical Engineering, Mechanical Engineering and Naval Architecture, University of Split, Split, Croatia\\
$^{34}$ Faculty of Nuclear Sciences and Physical Engineering, Czech Technical University in Prague, Prague, Czech Republic\\
$^{35}$ Faculty of Physics, Sofia University, Sofia, Bulgaria\\
$^{36}$ Faculty of Science, P.J.~\v{S}af\'{a}rik University, Ko\v{s}ice, Slovak Republic\\
$^{37}$ Faculty of Technology, Environmental and Social Sciences, Bergen, Norway\\
$^{38}$ Frankfurt Institute for Advanced Studies, Johann Wolfgang Goethe-Universit\"{a}t Frankfurt, Frankfurt, Germany\\
$^{39}$ Fudan University, Shanghai, China\\
$^{40}$ Gangneung-Wonju National University, Gangneung, Republic of Korea\\
$^{41}$ Gauhati University, Department of Physics, Guwahati, India\\
$^{42}$ Helmholtz-Institut f\"{u}r Strahlen- und Kernphysik, Rheinische Friedrich-Wilhelms-Universit\"{a}t Bonn, Bonn, Germany\\
$^{43}$ Helsinki Institute of Physics (HIP), Helsinki, Finland\\
$^{44}$ High Energy Physics Group,  Universidad Aut\'{o}noma de Puebla, Puebla, Mexico\\
$^{45}$ Horia Hulubei National Institute of Physics and Nuclear Engineering, Bucharest, Romania\\
$^{46}$ HUN-REN Wigner Research Centre for Physics, Budapest, Hungary\\
$^{47}$ Indian Institute of Technology Bombay (IIT), Mumbai, India\\
$^{48}$ Indian Institute of Technology Indore, Indore, India\\
$^{49}$ INFN, Laboratori Nazionali di Frascati, Frascati, Italy\\
$^{50}$ INFN, Sezione di Bari, Bari, Italy\\
$^{51}$ INFN, Sezione di Bologna, Bologna, Italy\\
$^{52}$ INFN, Sezione di Cagliari, Cagliari, Italy\\
$^{53}$ INFN, Sezione di Catania, Catania, Italy\\
$^{54}$ INFN, Sezione di Padova, Padova, Italy\\
$^{55}$ INFN, Sezione di Pavia, Pavia, Italy\\
$^{56}$ INFN, Sezione di Torino, Turin, Italy\\
$^{57}$ INFN, Sezione di Trieste, Trieste, Italy\\
$^{58}$ Inha University, Incheon, Republic of Korea\\
$^{59}$ Institute for Gravitational and Subatomic Physics (GRASP), Utrecht University/Nikhef, Utrecht, Netherlands\\
$^{60}$ Institute of Experimental Physics, Slovak Academy of Sciences, Ko\v{s}ice, Slovak Republic\\
$^{61}$ Institute of Physics, Homi Bhabha National Institute, Bhubaneswar, India\\
$^{62}$ Institute of Physics of the Czech Academy of Sciences, Prague, Czech Republic\\
$^{63}$ Institute of Space Science (ISS), Bucharest, Romania\\
$^{64}$ Institut f\"{u}r Kernphysik, Johann Wolfgang Goethe-Universit\"{a}t Frankfurt, Frankfurt, Germany\\
$^{65}$ Instituto de Ciencias Nucleares, Universidad Nacional Aut\'{o}noma de M\'{e}xico, Mexico City, Mexico\\
$^{66}$ Instituto de F\'{i}sica, Universidade Federal do Rio Grande do Sul (UFRGS), Porto Alegre, Brazil\\
$^{67}$ Instituto de F\'{\i}sica, Universidad Nacional Aut\'{o}noma de M\'{e}xico, Mexico City, Mexico\\
$^{68}$ iThemba LABS, National Research Foundation, Somerset West, South Africa\\
$^{69}$ Jeonbuk National University, Jeonju, Republic of Korea\\
$^{70}$ Johann-Wolfgang-Goethe Universit\"{a}t Frankfurt Institut f\"{u}r Informatik, Fachbereich Informatik und Mathematik, Frankfurt, Germany\\
$^{71}$ Korea Institute of Science and Technology Information, Daejeon, Republic of Korea\\
$^{72}$ KTO Karatay University, Konya, Turkey\\
$^{73}$ Laboratoire de Physique Subatomique et de Cosmologie, Universit\'{e} Grenoble-Alpes, CNRS-IN2P3, Grenoble, France\\
$^{74}$ Lawrence Berkeley National Laboratory, Berkeley, California, United States\\
$^{75}$ Lund University Department of Physics, Division of Particle Physics, Lund, Sweden\\
$^{76}$ Marietta Blau Institute, Vienna, Austria\\
$^{77}$ Nagasaki Institute of Applied Science, Nagasaki, Japan\\
$^{78}$ Nara Women{'}s University (NWU), Nara, Japan\\
$^{79}$ National and Kapodistrian University of Athens, School of Science, Department of Physics , Athens, Greece\\
$^{80}$ National Centre for Nuclear Research, Warsaw, Poland\\
$^{81}$ National Institute of Science Education and Research, Homi Bhabha National Institute, Jatni, India\\
$^{82}$ National Nuclear Research Center, Baku, Azerbaijan\\
$^{83}$ National Research and Innovation Agency - BRIN, Jakarta, Indonesia\\
$^{84}$ Niels Bohr Institute, University of Copenhagen, Copenhagen, Denmark\\
$^{85}$ Nikhef, National institute for subatomic physics, Amsterdam, Netherlands\\
$^{86}$ Nuclear Physics Group, STFC Daresbury Laboratory, Daresbury, United Kingdom\\
$^{87}$ Nuclear Physics Institute of the Czech Academy of Sciences, Husinec-\v{R}e\v{z}, Czech Republic\\
$^{88}$ Oak Ridge National Laboratory, Oak Ridge, Tennessee, United States\\
$^{89}$ Ohio State University, Columbus, Ohio, United States\\
$^{90}$ Physics department, Faculty of science, University of Zagreb, Zagreb, Croatia\\
$^{91}$ Physics Department, Panjab University, Chandigarh, India\\
$^{92}$ Physics Department, University of Jammu, Jammu, India\\
$^{93}$ Physics Program and International Institute for Sustainability with Knotted Chiral Meta Matter (WPI-SKCM$^{2}$), Hiroshima University, Hiroshima, Japan\\
$^{94}$ Physikalisches Institut, Eberhard-Karls-Universit\"{a}t T\"{u}bingen, T\"{u}bingen, Germany\\
$^{95}$ Physikalisches Institut, Ruprecht-Karls-Universit\"{a}t Heidelberg, Heidelberg, Germany\\
$^{96}$ Physik Department, Technische Universit\"{a}t M\"{u}nchen, Munich, Germany\\
$^{97}$ Politecnico di Bari and Sezione INFN, Bari, Italy\\
$^{98}$ Research Division and ExtreMe Matter Institute EMMI, GSI Helmholtzzentrum f\"ur Schwerionenforschung GmbH, Darmstadt, Germany\\
$^{99}$ Saga University, Saga, Japan\\
$^{100}$ Saha Institute of Nuclear Physics, Homi Bhabha National Institute, Kolkata, India\\
$^{101}$ School of Physics and Astronomy, University of Birmingham, Birmingham, United Kingdom\\
$^{102}$ Secci\'{o}n F\'{\i}sica, Departamento de Ciencias, Pontificia Universidad Cat\'{o}lica del Per\'{u}, Lima, Peru\\
$^{103}$ SUBATECH, IMT Atlantique, Nantes Universit\'{e}, CNRS-IN2P3, Nantes, France\\
$^{104}$ Sungkyunkwan University, Suwon City, Republic of Korea\\
$^{105}$ Suranaree University of Technology, Nakhon Ratchasima, Thailand\\
$^{106}$ Technical University of Ko\v{s}ice, Ko\v{s}ice, Slovak Republic\\
$^{107}$ The Henryk Niewodniczanski Institute of Nuclear Physics, Polish Academy of Sciences, Cracow, Poland\\
$^{108}$ The University of Texas at Austin, Austin, Texas, United States\\
$^{109}$ Universidad Aut\'{o}noma de Sinaloa, Culiac\'{a}n, Mexico\\
$^{110}$ Universidade de S\~{a}o Paulo (USP), S\~{a}o Paulo, Brazil\\
$^{111}$ Universidade Estadual de Campinas (UNICAMP), Campinas, Brazil\\
$^{112}$ Universidade Federal do ABC, Santo Andre, Brazil\\
$^{113}$ Universitatea Nationala de Stiinta si Tehnologie Politehnica Bucuresti, Bucharest, Romania\\
$^{114}$ University of Cape Town, Cape Town, South Africa\\
$^{115}$ University of Houston, Houston, Texas, United States\\
$^{116}$ University of Jyv\"{a}skyl\"{a}, Jyv\"{a}skyl\"{a}, Finland\\
$^{117}$ University of Kansas, Lawrence, Kansas, United States\\
$^{118}$ University of Liverpool, Liverpool, United Kingdom\\
$^{119}$ University of Science and Technology of China, Hefei, China\\
$^{120}$ University of South-Eastern Norway, Kongsberg, Norway\\
$^{121}$ University of Tennessee, Knoxville, Tennessee, United States\\
$^{122}$ University of the Witwatersrand, Johannesburg, South Africa\\
$^{123}$ University of Tokyo, Tokyo, Japan\\
$^{124}$ University of Tsukuba, Tsukuba, Japan\\
$^{125}$ Universit\"{a}t M\"{u}nster, Institut f\"{u}r Kernphysik, M\"{u}nster, Germany\\
$^{126}$ Universit\'{e} Clermont Auvergne, CNRS/IN2P3, LPC, Clermont-Ferrand, France\\
$^{127}$ Universit\'{e} de Lyon, CNRS/IN2P3, Institut de Physique des 2 Infinis de Lyon, Lyon, France\\
$^{128}$ Universit\'{e} de Strasbourg, CNRS, IPHC UMR 7178, F-67000 Strasbourg, France, Strasbourg, France\\
$^{129}$ Universit\'{e} Paris-Saclay, Centre d'Etudes de Saclay (CEA), IRFU, D\'{e}partment de Physique Nucl\'{e}aire (DPhN), Saclay, France\\
$^{130}$ Universit\'{e}  Paris-Saclay, CNRS/IN2P3, IJCLab, Orsay, France\\
$^{131}$ Universit\`{a} degli Studi di Foggia, Foggia, Italy\\
$^{132}$ Universit\`{a} del Piemonte Orientale, Vercelli, Italy\\
$^{133}$ Universit\`{a} di Brescia, Brescia, Italy\\
$^{134}$ Variable Energy Cyclotron Centre, Homi Bhabha National Institute, Kolkata, India\\
$^{135}$ Warsaw University of Technology, Warsaw, Poland\\
$^{136}$ Wayne State University, Detroit, Michigan, United States\\
$^{137}$ Yale University, New Haven, Connecticut, United States\\
$^{138}$ Yonsei University, Seoul, Republic of Korea\\
$^{139}$  Zentrum  f\"{u}r Technologie und Transfer (ZTT), Worms, Germany\\
$^{140}$ Affiliated with an institute formerly covered by a cooperation agreement with CERN\\
$^{141}$ Affiliated with an international laboratory covered by a cooperation agreement with CERN.\\

\end{flushleft} 